# Using Semantic Similarity for Input Topic Identification in Crawling-based Web Application Testing

Jun-Wei Lin and Farn Wang
Graduate Institute of Electrical Engineering
National Taiwan University, Taipei, Taiwan
{d01921014, farn}@ntu.edu.tw

## ABSTRACT
To automatically test web applications, crawling-based techniques are usually adopted to mine the behavior models, explore the state spaces or detect the violated invariants of the applications. However, in existing crawlers, rules for identifying the topics of input text fields, such as login ids, passwords, emails, dates and phone numbers, have to be manually configured. Moreover, the rules for one application are very often not suitable for another. In addition, when several rules conflict and match an input text field to more than one topics, it can be difficult to determine which rule suggests a better match. This paper presents a natural-language approach to automatically identify the topics of encountered input fields during crawling by semantically comparing their similarities with the input fields in labeled corpus. In our evaluation with 100 real-world forms, the proposed approach demonstrated comparable performance to the rule-based one. Our experiments also show that the accuracy of the rule-based approach can be improved by up to 19% when integrated with our approach.

## 1. INTRODUCTION
Web applications nowadays play important roles in our financial, social and other daily activities. Testing modern web applications is challenging because their behaviors are determined by the interactions among programs written in different languages and running concurrently in the front-end and the back-end. To avoid dealing with these complex interactions separately, test engineers treat the application as a black-box and abstract the DOMs (Document Object Models) presented to the end-user in the browser as states to model the behaviors of the application as a state transition diagram on which model-based testing can be conducted. Since manual state exploration is often labor-intensive and incomplete, crawling-based techniques [9, 10, 13, 14, 15, 24, 25, 27, 29] are introduced to systematically and automatically explore the state spaces of web applications. Although such techniques automate the testing of complicated web applications to a great extent, they are limited in valid input value generation. It is crucial for a crawler to provide valid input values to the application under test (AUT) because many web applications require specific input values to their input fields in order to access the pages and functions behind the current forms. To achieve proper coverage of the state space of the application, a user of existing crawlers needs to manually configure the rules for identifying the input topics in advance so as to feed appropriate input values at run time. For example, Figure 1 illustrates an input field requesting a first name, a value of the topic of *first_name*. To identify the topic of the input field, the values of its attributes such *id* and *name* have to be compared with a preset feature string, “firstName”, and an appropriate value can then be determined by the identified topic. Because input values in different topics such as email, URL and password are necessary for a web page requesting them, the manual configuration has to be repeated. Moreover, the rules for one application are likely not suitable for another, since the naming conventions for input fields in different web applications are diverse. Finally, it could be difficult to determine the topic of an encountered input field when it matches multiple rules for different topics. These drawbacks of the rule-based approach for input field topic identification has greatly limited the broad application of the existing crawling-based techniques.

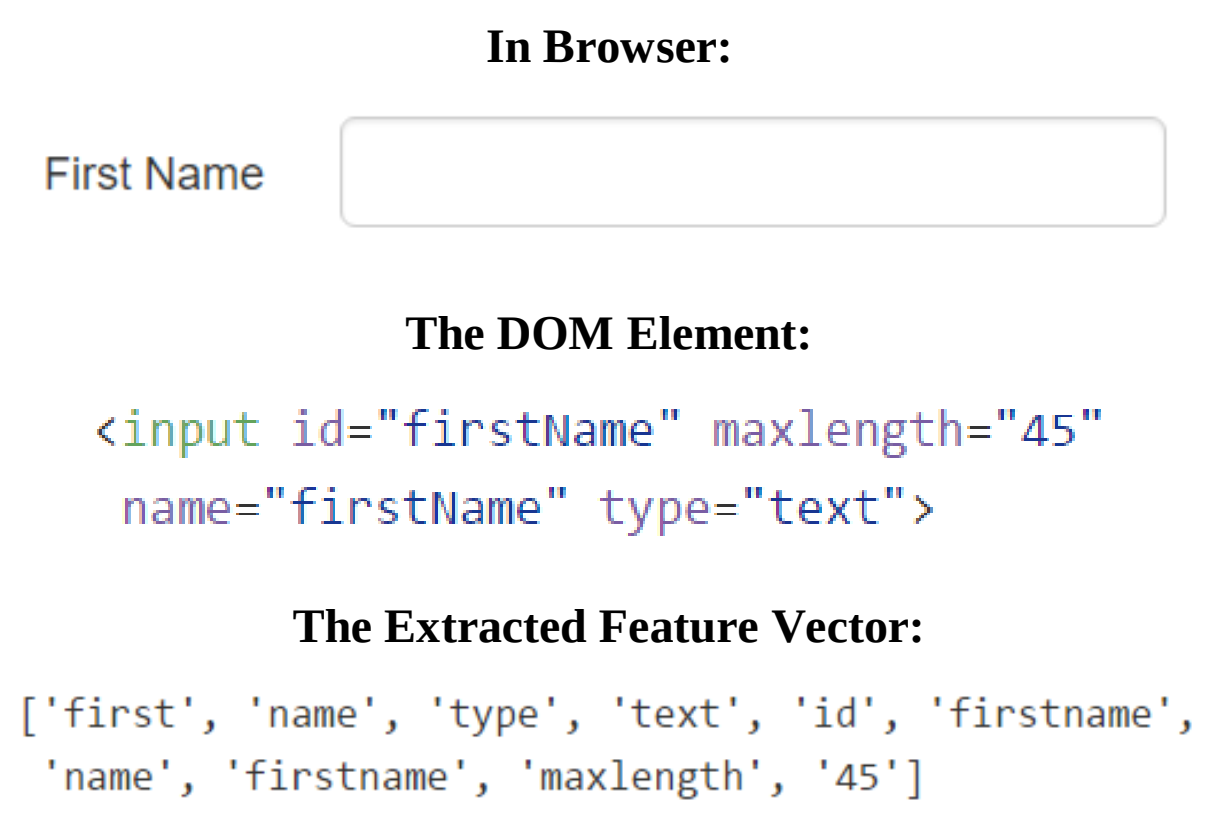

**Figure 1. An example input field asking for an first name, a value belonging to the topic of *first_name*.**

To address the issues of the rule-based approach for input topic identification in web application testing, several observations suggest the possibility of using natural-language techniques. First, in markup languages like HTML and XML, the words to describe the attributes of input fields such as *id*, *name*, *type*, and *maxlength* are extremely limited. As a result, unlike in a traditional natural-language task such as sentimental analysis which needs a large corpus, we could build a representative corpus of moderate size for the inference. Second, computer programs identify the topics of the input fields by looking at their DOM attributes, but human knows what to fill in by reading the corresponding labels or descriptions written in natural language. Finally, while the words and sentences used for input fields of the same topic may be different among web applications, they are usually semantically similar. For example, different websites may use “last name”, “surname”, “family name” or other related words to label and name the input fields taking the user’s last name. These observations formed the intuition of the proposed approach.

This paper presents a novel technique to automatically identify the topics of the input fields in web application testing. The proposed approach adopts techniques of natural language processing under supervised learning paradigm. First, in the training phase, for each encountered input field, we extract its feature vector consisting of the words in the DOM attributes and the nearest labels. An example feature vector is illustrated in Figure 1. We then build a training corpus with the feature vectors, and apply a series of transformations including Bag-of-words, Tf-idf (Term frequency with inverse document frequency) and LSI (Latent Semantic Indexing) [21] to the corpus, to represent the feature vectors with real numbers. These transformations discover relationships between words, and use them to describe the vectors in the corpus. The last stage of the training phase is labeling each feature vector in the corpus with a topic. Because after the transformations, the feature vectors are projected to a vector space in which each dimension of the space is related to a latent concept formed by the words in the corpus, the labeling process could be facilitated by a clustering heuristic explained in Section 3. Later in the inference phase, with the labeled corpus and vector space models, we can infer the topic of an unknown input field by figuring out its most similar vectors in the corpus under the same transformations. An appropriate input value for the recognized topic can then be selected from pre-established test data bank. We believe that the proposed method can relieve the burden of constructing rules for unexplored web applications, improve the effectiveness of input topic identification and enhance existing crawling-based techniques.

To evaluate the proposed approach, we conducted experiments with 100 real-world registration forms across different countries, and split them into training and testing data to validate the effectiveness of different identification approaches. The experimental results show that our approach performs comparably to the rule-based one as the proportion of training data increases, and the accuracy of the rule-based approach is significantly improved by up to 19% when integrated with the proposed natural-language technique. In addition, the experiments with real-world form submissions show that the proposed method outperforms the rule-based one in average.

The main contributions of this paper include:

- A novel technique using semantic similarity for input topic identification in web application testing to address the limitations of the rule-based approach in existing crawlers.
- An algorithm for introducing the corresponding labels or descriptions in addition to the DOM attributes of input fields to identify the topics.
- The implementation and evaluation of the proposed approach. Experiments with 100 real-world forms confirm the effectiveness of our approach. The source code and data of our experiment are also publicly available [3] to make the experiments reproducible.

## 2. BACKGROUND AND MOTIVATION

Today’s web applications interact responsively with the users by dynamically changing the DOMs using client-side JavaScript. To capture the behaviors of such applications, crawling-based technique plays a significant role [17] in automated web application testing [9, 10, 13, 14, 15, 24, 25, 27, 29]. The technique analyzes the data and models generated from dynamic exploration of the applications. Although exhaustive crawling can cause state explosion problem in most industrial web applications that have huge state spaces, the navigational diversity of the crawling is still important because we hope for deriving a test model with adequate functionality coverage [14]. However, achieving this diversity is challenging because many web applications require specific inputs to reach the hidden states behind input fields or forms [15]. For example, a web page querying for user profile data or valid URLs cannot be passed with random strings.

While there was a crawling technique ignoring text input [13], most existing crawlers [10, 24, 27] handle the input fields with randomly generated or user-specified data. To specify the data used in specific input fields, users have to provide feature strings (i.e., the strings to appear in the DOM attributes such as *id* or *name* of the input field) in rules to identify the topics. For example, if we want “Bob” for the input field with id “firstName”, we could add a line similar to the following when configuring the crawler:

```
input.field("first").setValue("Bob")
```

It is noteworthy that even though we do not specify the topic explicitly, we intuitively categorize the input fields with *id* or *name* containing string “first” as the topic of *first_name*, and then assign “Bob” as the value for the topic. Also, we use “first” instead of “firstName” because it can then be used to identify other possible input fields of the same topic such as the ones with id “first_name” or “s_user_first_nm”.

Rule-based identification for input topics is widely used in existing crawlers. Nevertheless, a couple of issues limit its application. The first is that the rules for one web application may not work for another. As a result, users may have to reconstruct or adjust their rules for new applications under test. For instance, Table 1 shows input fields collected from four real-world forms. The first input field contains two attributes, *id* and *name*, both with values “firstName”. To identify the input field and assign values used for it, the rule containing a feature string “first” is created to match the *id* or *name*. However, as illustrated, the rule derived from the first input field does not work for the second one which needs feature string "fn" in the rule. Moreover, both rules fail in identifying the third input field of the same topic, because the *id* and *name* look randomly generated. To address this issue, our approach takes the nearest labels or descriptions of a DOM element into consideration. The intuition is that the nearest labels are likely the texts about the input field for human to read, and if so, the texts for the same topics of input fields are usually semantically similar even in different websites. In fact, the third input field was successfully identified in our experiments as the first one was in training corpus.

**Table 1. Example input fields from real-world forms and rules for identifying them. Attributes except the *id* and *name* are removed for simplification. String: Feature string. TBD: To be discussed.**

| Input Field | Topic | String |
|---|---|---|
| <input id="firstName" name="firstName"> | first_name | "first" |
| <input id="aycreatefn" name="aycreatefn"> | first_name | "fn" |
| <input id="textfield-1028-inputEl" name="1023000000003015"> | first_name | (TBD) |
| <input id="permanenttel" name="permanenttel"> | phone | "tel" |
| <input id="aycreateln" name="aycreateln"> | last_name | "ln" |

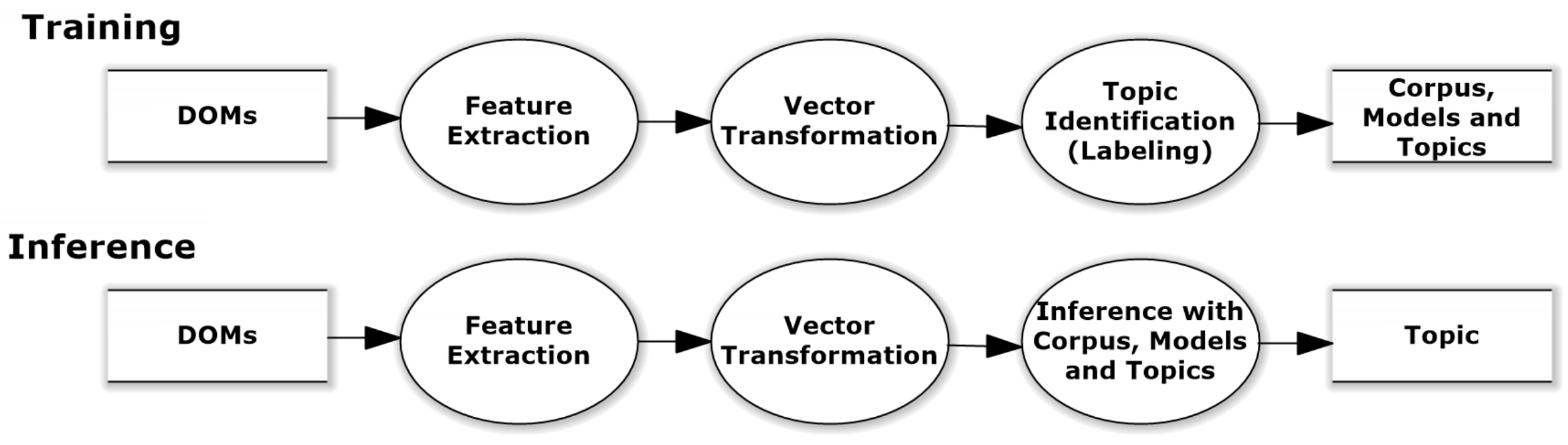

**Figure 2. The overview of the proposed approach.**

The second issue of the rule-based approach is that it is difficult to determine the topic if there are multiple candidates. For example, after setting the rules containing the feature string "tel" for the fourth input field and "ln" for the fifth input field in Table 1, the fifth input field will be categorized as *phone* and *last_name* simultaneously because both rules match the id value "aycreateln". In contrast, the proposed approach could resolve the ambiguity when considering the answer based on semantic similarity. The proposed method worked for the above example in our experiments and identified the topic correctly.

In unsupervised document analysis, vector transformations such as Tf-idf and LSI are algorithms that project a text corpus to a vector space by examining the word statistical co-occurrence patterns [21]. The concept behind Tf-idf is that the words appearing frequently in a document and infrequently in other documents could be used to uniquely represent the document. Furthermore, LSI is used to reduce the rank of a word-document matrix by applying Singular Value Decomposition [16], a mathematical technique in linear algebra. Each dimension of the dimension-reduced vector space hopefully represents a latent concept or topic in the texts. In this work, we apply these transformations to the feature vectors extracted from encountered input fields, and measure how similar two vectors are by calculating the cosine similarity (i.e., the cosine of the angle) between them [28]. The details are explained in Section 3.

# 3. APPROACH

The purpose of the proposed approach is to automatically identify the topics of encountered input fields in web application testing. Once the topics are identified, the corresponding values can then be selected from a pre-established databank or generated by data models such as smart profile [7]. Our approach includes two phases depicted in Figure 2. First, in the training phase, we extract feature vectors from collected input fields to build a training corpus, derive vector space models from the corpus and label each vector in the corpus by its topic. It must be noted that the labeling may need not to be performed one by one because the vectors are clustered by a heuristic proposed in Section 3.3. As a result, we may label a group of vectors as the same topic at a time. Second, with the artifacts from the training phase, the inference phase is fully automated. The feature vector of an encountered input field will be projected into the vector space constructed by training data, and the topic can then be determined by an algorithm based on its similarity to vectors in the training corpus. Each stage of the proposed approach is explained in the following subsections.

## 3.1 Feature Extraction

The first stage of the proposed approach is to extract the feature vector from an encountered input field which is expressed in a DOM element. A novelty of this paper is that we consider not only the attributes but also the nearby labels or descriptions of the DOM element in feature extraction. Algorithm 1 shows how it is achieved. First, we specify DOM attributes such as *id*, *name*, *placeholder* and *maxlength* which concerns input topic identification in an attribute list, and the matched attributes and their values of the DOM elements will be put into the feature vector (line 2 to 4). Moreover, to find the corresponding

**Algorithm 1** Extract Features

**Input:**
A DOM element $D$;
An attribute list $L$;
Maximum number of iterations $MAX$;
A tag list $T$;
**Output:**
A feature vector $F$ extracted from $D$;

```
1:  F = ∅;
2:  for each attribute ∈ D.attributes do
3:      if attribute ∈ L then
4:          F = F ∪ {attribute, D.getValue(attribute)}
5:  F = F ∪ findClosestLabels(D);

6:  function findClosestLabels(DOM element D, iteration=0)
7:      if iteration = MAX then
8:          return ∅
9:      else
10:         labels = ∅
11:         for each sibling ∈ D.siblings do
12:             for each tag_name ∈ T do
13:                 tags = sibling.find(tag_name)
14:                 if tags ≠ ∅ then
15:                     for each tag ∈ tags do
16:                         labels = labels ∪ tag.getText()
17:         if label ≠ ∅ then
18:             return labels;
19:         else
20:             return findClosestLabels(D.parent, iteration + 1);
```

descriptions, we search the siblings of the DOM element for tags such as *span* and *label* in a tag list and put the texts enclosed by the tags into the feature vector (line 11 to 18). If no such tags found, the search will continue on the DOM's parent recursively for several times (line 20). In addition, we perform a couple of normalizations such as special character filtering and lowercase conversion to the words in the extracted feature vector. An example of an extracted feature vector is shown in Figure 3. For the input field in Figure 3, the feature vector was first constructed with its attributes and values: "type", "text", "id", "firstname", "name", "firstname", "maxlength" and "45" (line 2 to 4). Then in *findClosestLabels()*, because the input element has no siblings, the algorithm searched siblings of its parent (line 20). In the second iteration of the search, a label with text "First Name" was found, and the words "first" and "name" were put into the feature vector. The same process is also adopted in the inference phase.

**DOM:**

```
<div class="control-group">
 <label for="firstName" class="control-label">
  First Name
 </label>
 <div class="controls">
  <input id="firstName" name="firstName" maxlength="45"
   type="text">
 </div>
</div>
```

**The Extracted Feature Vector:**

```
['first', 'name', 'type', 'text', 'id', 'firstname',
 'name', 'firstname', 'maxlength', '45']
```

**Figure 3. An example of an input field and the extracted feature vector.**

## 3.2 Vector Transformation

After all input fields for training are represented as feature vectors in a corpus, three transformations are applied to the vectors sequentially: Bag-of-words, Tf-idf and LSI [21]. These transformations convert the vectors from words to real numbers, and project the vectors to a vector space in which each dimension of the space hopefully represents a latent topic consisting of the words in the corpus. As a result, we could cluster the input fields according to their calculated weights in each dimension in the latent space. In the following paragraphs we use document and feature vector interchangeably because they are identical in the context of this section.

First, bag-of-words transformation is used to represent each document in natural language in a corpus as an integer vector based on its word counts. In general, the dimension of the integer vector is the number of distinct words in the corpus. For example, if a document is "John likes cat, and Mary likes cat, too" in a corpus, after filtering some common words with high frequency (called *stopwords* in natural language processing) such as "and" and "too", the document could be represented as [1, 2, 2, 1, 0, 0, …, 0], which means "John" appeared once, "likes" and "cat" appeared twice, "Mary" appeared once and all other words didn't appear in it. Bag-of-words transformation is a simplified representation because it disregards grammar and word order in documents. Fortunately, for DOM elements of input fields in web applications we do not care about these two properties either.

Second, Tf-idf transformation converts the bag-of-word integer counts to real-value weights. Intuitively, if a word appears frequently in a document and infrequently in all other documents, the word could uniquely represent the document. Tf-idf assigns weights to words in documents based on this intuition. A common weighting scheme is:

$$f_{t,d} \times log\frac{N}{n_t}$$

$f_{t,d}$ here is the frequency of the word $t$ in document $d$, $N$ is the number of all documents and $n_t$ is the number of documents in which $t$ appears.

Finally, LSI transformation tries to deal with the problem that different words used in the same context may have similar meanings. The transformation reduces the dimension of the vector space constructed by the words and documents in a corpus using a mathematical technique called Singular Value Decomposition [16]. Each dimension of the rank-reduced vector space hopefully represents a latent concept or topic contained in documents. An example of feature vectors in words and the converted feature vectors in real numbers after the above three transformations is shown in Figure 4. To make the following explanation clearer, we

**A Real-world Form in Browser**

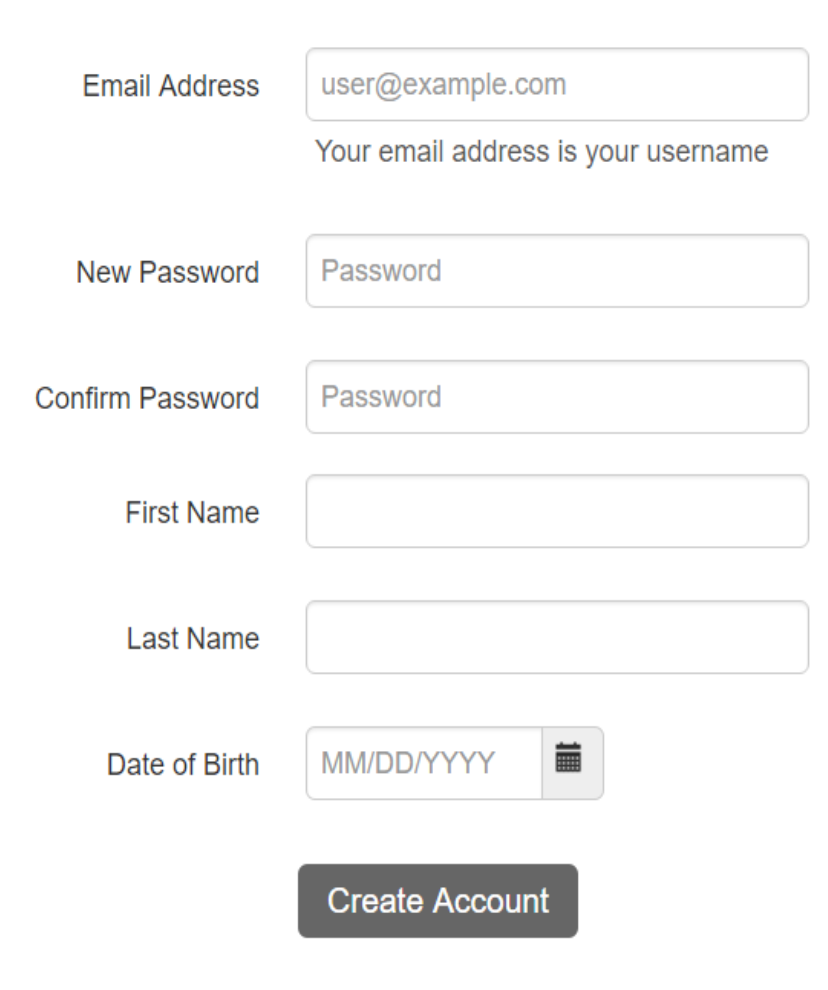

**Feature Vector and the Converted Representation after the Transformations**

| |
|---|
| ['your', 'email', 'address', 'is', 'your', 'username', 'name', 'email', 'type', 'text', 'placeholder', 'user', 'example', 'com', 'id', 'email', 'maxlength', '100']<br>[(0, 0.017518), (1, 0.021639), (2, -0.005462), (3, -0.00084), (5, 0.711760), …, (11, -0.000567)] |
| ['new', 'password', 'name', 'password', 'type', 'password', 'placeholder', 'password', 'id', 'password', 'maxlength', '80']<br>[(0, 0.913020), (1, -0.094889), (2, -0.000177), (3, 5.05e$^{-6}$), (5, -0.013603), …, (11, -0.396488)] |
| ['confirm', 'password', 'type', 'password', 'placeholder', 'password', 'id', 'confirmpassword', 'name', 'confirmpassword', 'maxlength', '80']<br>[(0, 0.913239), (1, -0.092972), (2, -0.000165), (3, 4.56e$^{-6}$), (5, -0.011448), …, (11, 0.396509)] |
| ['first', 'name', 'type', 'text', 'id', 'firstname', 'name', 'firstname', 'maxlength', '45']<br>[(0, 0.000445), (1, 0.005451), (2, -0.424814), (3, 0.300439), (4, -0.573656), …, (11, 1.77e$^{-6}$)] |
| ['last', 'name', 'type', 'text', 'id', 'lastname', 'name', 'lastname', 'maxlength', '45']<br>[(0, 0.000445), (1, 0.005451), (2, -0.424814), (3, 0.300439), (4, 0.573656), …, (11, 1.77e$^{-6}$)] |
| ['date', 'of', 'birth', 'name', 'dateofbirth', 'type', 'text', 'placeholder', 'mm', 'dd', 'yyyy', 'id', 'date', 'input', 'maxlength', '10']<br>[(0, 0.020378), (1, 0.025152), (2, -0.006330), (3, -0.000969), (5, 0.705647), …, (11, -0.000663)] |

**Figure 4. An example of feature vectors and the converted representations after applying the transformations.**

use sparse representation for the converted feature vectors. That is, each tuple in a vector is a dimension index followed by the calculated weight of the dimension, and the weight ranges from -1 to 1. The converted feature vector of the first input field taking email is:

$$\begin{bmatrix} (0, 0.017518), \\ (1, 0.021639), \\ (2, -0.005462), \\ (3, -0.00084), \\ (5, 0.711760), \\ \cdots, \\ (11, -0.000567) \end{bmatrix}$$

which means the weight of the input field in concept 0 is 0.0017518, the weight in concept 1 is 0.021639, the weight in concept 2 is -0.005462 and so on. It must be noted that the concepts here are latent concepts justified on the mathematical level and probably have no interpretable meaning in natural language. They are not the topics defined for labeling and testing. In this example the number of concepts or dimensions is twelve, and it varies with the words and size of a corpus.

## 3.3 Topic Identification and Labeling

At this stage, each input field in the training corpus is labeled by its topic, and we can take advantage of the results from previous stages to facilitate the labeling process. First, conceptually similar input fields are expected to be close in the latent vector space. For example, the converted vectors of the second and the third input fields in Figure 4 are close:

$$\begin{bmatrix} (0, 0.913020), \\ (1, -0.094889), \\ (2, -0.000177), \\ (3, 5.05e^{-6}), \\ (5, -0.013603), \\ \cdots, \\ (11, -0.396488) \end{bmatrix}, \begin{bmatrix} (0, 0.913239), \\ (1, -0.092972), \\ (2, -0.000165), \\ (3, 4.56e^{-6}), \\ (5, -0.011448), \\ \cdots, \\ (11, 0.396509) \end{bmatrix}$$

and both the input fields should be labeled as the topic of *password* because they all take passwords. In addition, we notice that if we want to pick a latent concept to represent the above two vectors, the concept 0 seems most appropriate since the weights in this dimension are maximal over all dimensions of the two vectors, respectively. As a result, we developed a heuristic to quickly map each input fields to a latent concept, and a user can choose to label the input fields belonging to the same latent concept at a time. An input field will be mapped to the latent concept in which the absolute weight of the converted vector is maximal. For instance, the second and the third input fields in Figure 4 will be mapped to the latent concept 0 because their maximal weights of the converted vectors are both in dimension 0 (0.913020 and 0.913239). Moreover, the converted vectors of the fourth input field for first name and the fifth for last name in Figure 4 are:

$$\begin{bmatrix} (0, 0.000445), \\ (1, 0.005451), \\ (2, -0.424814), \\ (3, 0.300439), \\ (4, -0.573656), \\ \cdots, \\ (11, 1.77e^{-6}) \end{bmatrix}, \begin{bmatrix} (0, 0.000445), \\ (1, 0.005451), \\ (2, -0.424814), \\ (3, 0.300439), \\ (4, 0.573656), \\ \cdots, \\ (11, 1.77e^{-6}) \end{bmatrix}$$

These two input fields will be mapped to the latent concept 4 because their absolute weights of the converted vectors in dimension 4 are both 0.573656 and are maximal. User can label a cluster of input fields provided by the heuristic with a topic for inference, or choose to label some of them separately.

## 3.4 Inference with the Models and Topics

In the inference phase, for an encountered input field, we first extract its feature vector with the same process described in Section 3.1, and then transform and project the vector to the same latent space with the vector models derived in the training phase. To calculate the similarity between two vectors, we adopt cosine similarity, i.e., the cosine of the angle, because it is reported a good measure in information retrieval [28]. The cosine similarity of two vectors A and B is:

$$\frac{A \cdot B}{\|A\|\|B\|}$$

Algorithm 2 describes how we determine the input topic based on its cosine similarities to training data. First, the topic of the vector in the latent space most similar to the encountered one will be selected (line 3 to 4). If the difference of the similarities between the top 5 most similar vectors is less than a threshold, the topic will be determined by a voting process within the top 5 vectors (line 5 to 8). If there are multiple candidates after the vote, a random choice will be made (line 10). The voting process provides a chance to correctly infer the topic when there are multiple vectors with close similarity scores. For example, Table 2 depicts that the inferred topic is mistaken when only the most similar vector is considered, but there is a chance to correct the mistake since the voting process may guess the right topic in random choice from *last_name* and *password*.

**Algorithm 2** Infer Topic

**Input:**
A DOM element $D$;
A vector space model $V$;
A mapping table $MT$ for mapping feature vectors in training corpus to the topics;
A threshold $THRESHOLD$;
**Output:**
The topic of $D$;

```
1:  vec = extractFeatures(D);
2:  vec = mapToSpace(vec, V);
3:  list = findMostSimilarVectors(vec, V);        ▷ Return top 5 vectors in V
    ranked by cosine similarities to vec
4:  inferredTopic = vecToTopic(list[0], MT);
5:  if (Similarity(list[4]) − Similarity(list[0])) < THRESHOLD then
6:      candidates = occurMostTimes(list, MT);     ▷ Return topics occured
    most times in the list
7:      if |candidates| = 1 then
8:          inferredTopic = candidates[0];
9:      else
10:         inferredTopic = randomlyChoose(candidates);
11: return inferredTopic;
```

## 3.5 Integration with the Rule-based Approach

The issues of the rule-based approach mentioned in Section 2 can be addressed by integrating with the proposed approach. Algorithm 3 provides the details. First, for the input fields not identified by the rule-based method, we output the answer found by our technique (line 4 to 5). Second, for the input fields matching multiple rules for different topics, we select the answer with the help of the natural-language technique (line 6 to 11). Specifically, if the natural-language answer appears in the candidates, the answer will be selected (line 8 to 9), or a random choice will be made among all candidates including the natural-language one (line 11). In Section 5, we evaluated the effectiveness of the integration.

**Table 2. An example feature vector and its top 5 most similar vectors.**

| Similarity | Vector | Topic |
|---|---|---|
| (N/A) | ['type', 'text', 'name', 'psurname', 'maxlength', '30'] (To be inferred) | last name |
| 0.998230 (Most similar) | ['type', 'text', 'name', 'fname', 'maxlength', '30'] | first name |
| 0.997782 | ['type', 'text', 'name', 'lname', 'maxlength', '30'] | last name |
| 0.996292 | ['according', 'to', 'official', 'document', 'name', 'familyname', 'id', 'familyname', 'type', 'text', 'maxlength', '30'] | last name |
| 0.929636 | ['name', 'conf', 'pword', 'type', password', 'maxlength', '30'] | password |
| 0.921988 | ['name', 'pword', 'type', 'password', 'maxlength', '30'] | password |

**Algorithm 3** Improve the Rule-based Method with the Proposed Approach

**Input:**
A DOM element $D$;
**Output:**
The topic of $D$;

```
topics = inferWithRules(D);
if |topics| = 1 then
    inferredTopic = topics[0];
else if topics = ∅ then                                   ▷ No match
    return inferWithNL(D);                                ▷ Refer to Algorithm 2
else                                                      ▷ Multiple candidates
    topic_NL = inferWithNL(D)
    if topic_NL ∈ topics then
        inferredTopic = topic_NL
    else
        inferredTopic = randomlyChoose(topics ∪ topic_NL)
return inferredTopic;
```

# 4. IMPLEMENTATION

We implemented the proposed method with Python 2.7. A Python library, gensim [26], is used for vector space related operations such as vector transformation and similarity calculation. Interaction with web applications is supported by Selenium Webdriver [2], and BeautifulSoup [1] is used to parse and manipulate DOMs.

# 5. EVALUATION

To assess the efficacy of the proposed approach, we conducted two controlled experiments with 100 real-world forms. In the first experiment, we analyzed and labeled the input fields in the forms, used some forms to build training corpora and derive rules, and evaluated the performances of the proposed and rule-based approaches. In the second experiment, 35 simple forms from the 100 forms were actually tested with identified input topics and corresponding values with different methods. Two research questions were addressed:

**Q1.** What is the effectiveness of the proposed approach comparing with the rule-based one? How much training data is needed?

**Q2.** Can the proposed approach be used to improve the rule-based one?

Our experimental data along with the implementation can be accessed publicly [3].

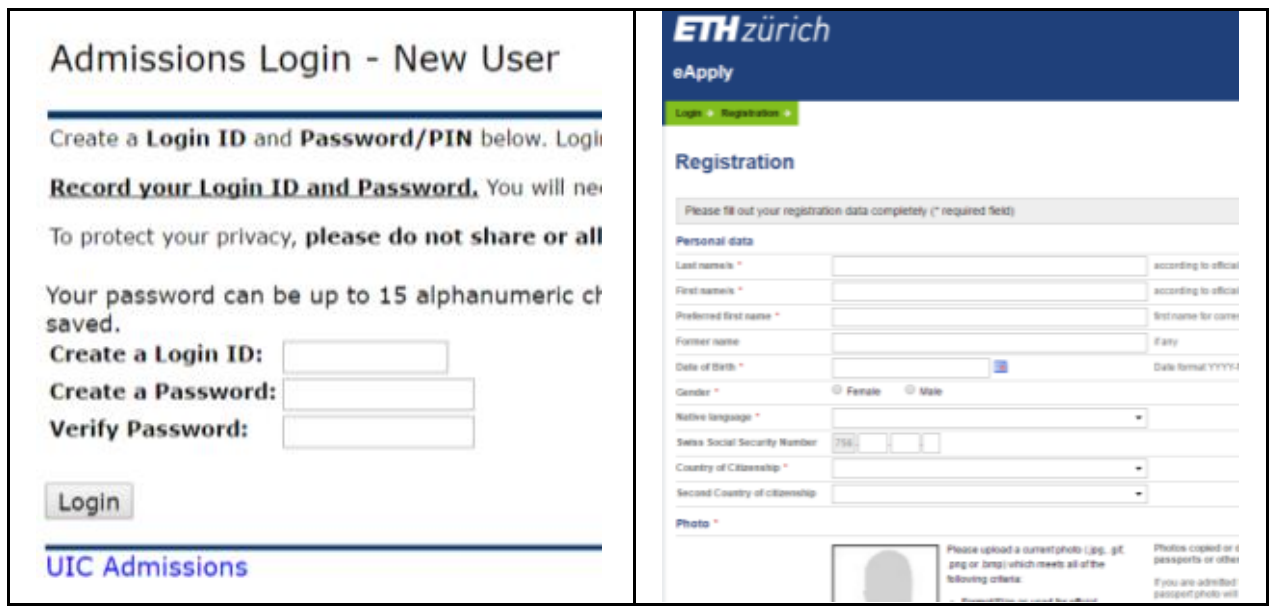

**Figure 5. Screenshots of two subject forms containing three and twenty-five input fields, respectively.**

## 5.1 Subject Forms

We collected 100 graduate program registration forms across 9 countries in the world (the complete list is provided in appendix), and two examples of the forms are shown in Figure 5. There are totally 958 input fields in the forms, ranging from two to fifty-eight for each form, and 62 input topics such as *password*, *email*, *first_name* and *zipcode* are labeled. Table 3 shows the labeled topics and the number of input fields for each topic. These topics have to be distinguished from each other to pass the forms. For example, several date-related topics with different formats such as *date-mm/dd/yyyy*, *date-mm/yyyy* and *year-yyyy* are defined for input fields in different forms taking date information. We choose registration forms as subject data for several reasons. First, they usually contain many different topics of input fields such as user profile, date or URL, which is appropriate for our evaluation. Second, the application states behind forms are important because they take information from the users and then interact with them. However, the states are usually difficult to be reached using existing crawlers with random inputs. In addition, we want to evaluate the effectiveness of the methods on inferring unknown forms with training data in the same category. As a result, we collected only registration forms for the experiments. It is worth noting that even if we use forms as subjects and many input fields are presented within forms, the proposed technique is for all input fields in web applications.

**Table 3. Labeled input topics and number of input fields for each topic in the experiment.**

| Topic | # | Topic | # | Topic | # | Topic | # | Topic | # | Topic | # |
|---|---|---|---|---|---|---|---|---|---|---|---|
| password | 188 | street-line-2 | 13 | year-yyyy | 4 | street | 2 | promo_code | 1 | date-yyyy-mm-dd | 1 |
| email | 151 | street-line-1 | 13 | phone-middle-postfix | 3 | portal | 2 | data_collection | 1 | unknown_hidden | 1 |
| last_name | 105 | secure_ans | 11 | ssn | 3 | address-additional | 2 | date-yyyymmdd | 1 | ssn-postfix | 1 |
| first_name | 105 | special_id | 9 | phone-postfix | 3 | date | 2 | mname-1-char | 1 | user_status | 1 |
| username | 48 | search_term | 9 | street-line-3 | 3 | default_local_node | 2 | employer | 1 | visa_number | 1 |
| middle_name | 46 | full_name | 8 | phone-middle | 3 | ssn-swiss-middle-4 | 2 | validation_action | 1 | ssn-prefix | 1 |
| phone | 46 | phone-prefix | 8 | local_node | 2 | url | 2 | digit-16 | 1 | complex_pwd | 1 |
| date-mm/dd/yyyy | 43 | captcha | 7 | suffix | 2 | db_name | 2 | ssn-middle | 1 | Total | 985 |
| zipcode | 41 | short_pwd | 5 | alien_number | 2 | ssn-swiss-prefix-3 | 1 | secure_q | 1 | | |
| date-mm/yyyy | 28 | state | 5 | date-dd/mm/yyyy | 2 | date-dd | 1 | job_title | 1 | | |
| city | 25 | school_name | 5 | room_number | 2 | phone-ext | 1 | ssn-swiss-postfix-2 | 1 | | |

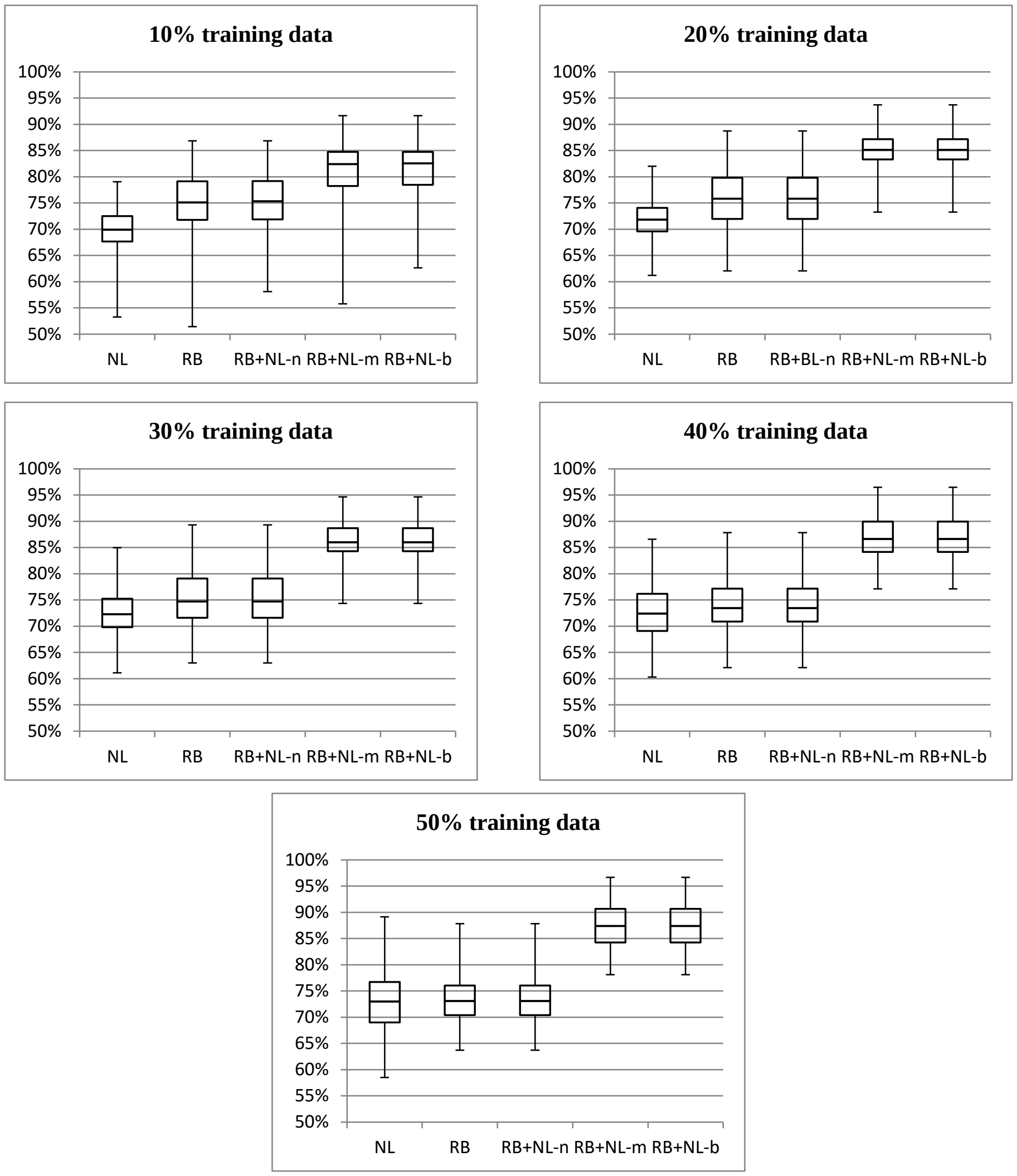

**Figure 6. Variance of the accuracy across the methods over.**

## 5.2 Experimental Setup

In the first experiment, to understand how the proportion of training data affects the performances of the methods under evaluation, we randomly chose 10%, 20%, 30%, 40% and 50% of the subject forms as training data, respectively. We then derived artifacts such as labeled corpus, vector models (for the proposed approach) and rules (for the rule-based approach) from the input fields of the training forms, and used the artifacts to infer the input fields in the remaining forms. Finally the inference accuracy was calculated to show the percentage of correctly identified input fields in the remaining forms. The experiments were repeated 1000 times. Five methods were evaluated in the first experiments: (1) NL, the proposed natural-language approach. (2) RB, the manual, rule-based approach. (3) RB+NL-n (no-match), using the NL approach to identify input fields not recognized by the RB approach, as discussed in Section 3.5 (4) RB+NL-m (multiple), using the NL approach to help identify input fields with multiple candidates by the RB approach, as discussed in Section 3.5. (5) RB+NL-b (both), using both (3) and (4).

In the second experiment, to evaluate the methods on real-world applications, 35 simple forms containing no elements such as radio buttons and dropdown lists but only input fields from the 100 forms were selected. These simple forms can only be successfully submitted with appropriate input values, and therefore are appropriate subjects for evaluating the methods discussed in this paper. Seven (i.e., 20%) of the forms were randomly picked as training data, and used to infer the topics of the input fields in the remaining 28 forms. We then submitted the forms with values corresponding to the identified topics from a data pool, and judged the testing results by test oracles collected from manually submissions. It is noteworthy that the forms could be passed through with values of incorrectly inferred topics, so this experiment provides another perspective to the effectiveness of the methods under evaluation. To avoid overwhelming the subject websites with experimental data, we only repeated the experiment 10 times. Also, we make sure that the values used in each trial are different and new. That is, submission failures won't be caused by duplicated data. Four methods were evaluated in the second experiment: (1) Random, submission with random 8-char strings. (2) NL. (3) RB+NL-n. (4) RB+NL-m.

## 5.3 Results and Discussion

### *5.3.1 Experiments with topic data*

Table 4 shows the average accuracies that each method achieved when the considered percentages are used as training data. First, with only 10% of the subject forms as training data, the proposed approach and the rule-based one both performed well in inferring the input fields of the rest 90%, with average accuracy 69.84% and 75.11%, respectively. Moreover, the average accuracy of the proposed approach increases with the percentage used as training data, but the performance of the rule-based approach slightly decreases as the proportion of training data increases. As a result, the proposed approach performs comparably to the rule-based one with 50% as training data. Second, while RB+NL-n achieve some improvement with 10% as training data, in general the improvement by identifying the no-match elements with the proposed approach is not significant. On the other hand, using the proposed approach to help pick the correct topic from multiple candidates by the rule-based approach can greatly improve the accuracy. For example, with 50% as training data, RB+NL-m outperforms average accuracy of RB by 19% (13.98% increase).

**Table 4. Average accuracies achieved by different methods when the considered percentages are used as training data.**

| % training | NL | RB | RB+ NL-n | RB+ NL-m | RB+ NL-b |
|---|---|---|---|---|---|
| 10% | 69.84% | 75.11% | 75.25% | 81.56% | 81.70% |
| 20% | 71.95% | 75.83% | 75.83% | 85.03% | 85.03% |
| 30% | 72.53% | 75.39% | 75.39% | 86.31% | 86.31% |
| 40% | 72.70% | 74.26% | 74.26% | 87.00% | 87.00% |
| 50% | 72.94% | 73.42% | 73.42% | 87.40% | 87.40% |

To determine whether the improvements we observed in average accuracies are statistically significant, we conducted a *t-test for matched pairs*[1] [18] for the NL, RB and RB+NL-b approaches. That is, the accuracies of these three methods in each trial are considered matched pairs to each other. We assume that there is no difference in the average accuracies by NL and RB, NL and RB+NL-b and RB and RB+NL-b, respectively (the *null hypotheses*). If the computed p-value is less than 0.05 (the *significance level*), statistical practitioners often infer that the null hypothesis is false. Table 5 shows the *p-values* computed for these three methods in our experiments. It indicates that the observed differences between these three methods are statistically significant. In addition, Figure 6 reports the variances of the accuracies of the 1000 runs for each method. It also demonstrates the differences in accuracy among techniques under evaluation.

To further understand the experimental results, we investigated the average number of no-match elements and elements with multiple candidate topics when adopting the rule-based approach. We also calculated the number of inferred elements. As Table 6 shows, with larger proportion of training data, there are less no-match and more multiple-topic elements. The observation is reasonable because with more training data introduced, the rule set derived from them is larger, and more input elements are likely to match multiple rules for different topics. We believe that the observation contributes to both the decreased accuracy of RB on increased training data and the improvement of RB+NL-m comparing with RB. In addition, the improvement of RB+NL-n comparing with RB is not significant, which could result from two reasons. First, the average number of no-match elements is not high. Second, from Table 2 we can see that many topics of the input fields appear only a few times. In fact, 12.4% of total topics appear less than 10 times, and identifying these topics may be difficult because the input fields are not included in the training data.

**Table 5. Computed p-values of t-test for matched pairs.**

| % training | NL & RB | NL & RB+NL-b | RB & RB+NL-b |
|---|---|---|---|
| 10% | 0.000000 | 0.000000 | 0.000000 |
| 20% | 0.000000 | 0.000000 | 0.000000 |
| 30% | 0.000000 | 0.000000 | 0.000000 |
| 40% | 0.000000 | 0.000000 | 0.000000 |
| 50% | 0.004284 | 0.000000 | 0.000000 |

**Table 6. Average number of inferred, no-match and multiple-topic elements in the experiments with rule-based approach.**

| % training | # inferred | # no-matches | | # multiple-topic | |
|---|---|---|---|---|---|
| | | # | % | # | % |
| 10% | 887.32 | 120.13 | 13.46% | 142.16 | 16.06% |
| 20% | 789.24 | 74.18 | 9.31% | 176.92 | 22.46% |
| 30% | 690.92 | 54.54 | 7.78% | 180.16 | 26.13% |
| 40% | 591.17 | 42.49 | 7.04% | 172.10 | 29.18% |
| 50% | 492.35 | 33.40 | 6.59% | 153.65 | 31.31% |

**Table 7. Number of passed forms with different methods.**

| Trial # | Random | NL | RB | RB+NL-n | RB+NL-m |
|---|---|---|---|---|---|
| 1 | 2 | 20 | 8 | 7 | 23 |
| 2 | 2 | 22 | 24 | 24 | 25 |
| 3 | 2 | 24 | 25 | 25 | 25 |
| 4 | 2 | 22 | 25 | 24 | 25 |
| 5 | 1 | 24 | 3 | 3 | 24 |
| 6 | 2 | 21 | 25 | 25 | 25 |
| 7 | 1 | 23 | 7 | 7 | 26 |
| 8 | 2 | 18 | 23 | 21 | 23 |
| 9 | 1 | 23 | 12 | 12 | 24 |
| 10 | 2 | 20 | 23 | 21 | 23 |
| Avg. | 1.70 | 21.70 | 17.50 | 16.90 | 24.30 |

### *5.3.2 Experiments with Real Forms*

The results of the second experiments are shown in Table 7. In these 10 trials, we can see that the random generated values are not helpful in passing the forms and only 1.7 of 28 forms were passed in average. On the other hand, the proposed method (NL) performs better than the rule-based one (RB) in average and in some cases, and the effectiveness is relatively stable in terms of number of passed forms. Moreover, one may notice that RB+NL-n performs worse than the original RB. By investigating the passed forms in RB but failed in RB+NL-n, we found that the input fields which no rule matched were assigned random values in RB, but in RB+NL-n the fields were assigned specific values such as emails or passwords which contain special characters. The observation suggests that incorrectly identified input topics may undermine the exploring capability of a crawler. Finally, the results of RB+NL-m in Table 7 show that the rule-based approach can be significantly improved with the proposed approach, which is consistent with the results of the first experiment.

[1] A *t-test for matched pairs* is a statistical method used to infer the statistical significance of the difference between the means of two populations, given samples where each observation in one sample is logically matched with an observation in the other sample. The testing procedure begins with a null hypothesis that assumes the population means are identical, and then computes a p-value from the paired data samples. Should the p-value be less than a selected significance level, the null hypothesis would be rejected.

### *5.3.3 Threats to Validity*

The implementation of the proposed approach could affect the validity of results. To ensure the correctness, we adopted mature and open-sourced libraries such as gensim [26] , BeautifulSoup [1] and Selenium Webdriver [2] in key steps of our implementation. In addition, the subject forms selected and the setup of the evaluation such as the defined topics and derived rules might affect generality of the results. To make the experimental data representative, we collected real-world forms across 9 countries in the world. We also open our source code and data to the public [3] for review and replication.

## 6. RELATED WORK

Crawling-based techniques for modern web applications have been studied [10, 13, 22, 24, 27] and adopted [9, 14, 15, 25, 29] in automated web application testing. Duda et al. [13] proposed algorithms to crawl AJAX-based web applications and index the states. Similarly, a tool developed by Mesbah et al. [24] called *Crawjax* tries to infer a finite state machine of an AJAX-based web application through dynamically analyzing the changes of the DOM contents. The tool is also used for detecting and reporting violated invariants [25] and cross-browser incompatibilities [9] in web applications. Schur et al. [27] presented a crawler called *ProCrawl* to extract abstract behavior models from multi-user web applications, focusing on building a model close to business logic. A crawler developed by Dallmeier et al. [10], *WebMate*, can autonomously explore a web application, and use existing test cases as an exploration base. Marchetto et al. [22] extracts a finite state machine from existing execution traces of web applications, and generates test cases consisting of dependent event sequences. In addition, Fard et al. [15] combined the knowledge inferred from manual test suites with automatic crawling in test case generation for web applications. Thummalapenta et al. [29] presented a technique to confine the number of a web application's GUI states explored by a crawler with existing business rules. A couple of metrics such as JavaScript code coverage, path diversity and DOM diversity are also proposed to evaluate the test model derived by a crawler [14]. However, none of these studies considers leveraging semantic similarity in input value handling as our work does.

Studies on GUI ripping for testing purpose [5, 23] and automatic black-box testing on mobile applications [4, 20] are also related to our work in terms of how they explore the interfaces of the applications and derive test models with dynamic analysis. As a result, the proposed technique could be applied in these contexts.

With respect to using latent topic models in software testing and debugging, Andrzejewski et al. [6] approached debugging using a variant of Latent-Dirichlet Allocation [8] to identify weak bug topics from strong interference. Latent-Dirichlet Allocation was also adopted by Lukins et al. [19] on a developer's input such as a bug report to localize faults statistically. Later, DiGiuseppe and Jones [11, 12] adopted natural-language techniques such as feature extraction and Tf-idf in fault description and clustering. To the best of our knowledge, this paper is the first to apply latent topic models to test input generation for web applications.

## 7. CONCLUSION

In this paper, we proposed a natural-language technique for input topic identification in web application testing. With vector space models and topics derived from training corpus, the topics of encountered input fields are inferred based on their semantic similarities to vectors in training corpus. For the recognized topic an appropriate input value can then be determined. The proposed approach addresses the issues of the rule-based one in existing crawlers to make them more applicable. Our evaluation shows that the proposed method performs comparably to the rule-based method, and the accuracy of the rule-based approach can be greatly improved with the proposed technique. In the experiment with real forms, the proposed technique performs better than the rule-based one in average and in some cases, and the effectiveness is relatively stable. In the future, we plan to conduct experiments with more data, and explore the possibility of applying our technique in different contexts such as clickable identification and state abstraction. Moreover, the proposed feature extraction algorithm could be improved with text containing more information about input fields such as comments.

## APPENDIX: The Real-world Forms Used in the Experiments

| School | Country | URL | School | Country | URL |
|---|---|---|---|---|---|
| Brigham Young University | US | https://tinyurl.com/jp2p7a5 | The University of Warwick | GB | https://tinyurl.com/z3y4ma3 |
| Brown University | US | https://tinyurl.com/htchnha | Tsinghua University | CN | https://tinyurl.com/guoy5mr |
| California Institute of Technology | US | https://tinyurl.com/zznjb8s | Tufts University | US | https://tinyurl.com/j7qcq4v |
| Carnegie Mellon University | US | https://tinyurl.com/jwzgmpx | University of Arizona | US | https://tinyurl.com/hvbh6du |
| Colorado State University | US | https://tinyurl.com/gw5dxut | University of Calgary | CA | https://tinyurl.com/jcgadpu |
| Columbia University | US | https://tinyurl.com/jruuvf5 | University of California, Berkeley | US | http://tinyurl.com/zmpyq4t |
| Cornell University | US | http://tinyurl.com/hejvyya | University of California, Davis | US | https://tinyurl.com/zsjcsos |
| Dartmouth College | US | https://tinyurl.com/ztg5xk5 | University of California, Irvine | US | https://tinyurl.com/hutg5fd |
| Drexel University | US | https://tinyurl.com/haw6q2x | University of California, Los Angeles | US | https://tinyurl.com/gs4377s |
| Duke University | US | https://tinyurl.com/qfrhlgv | University of California, Riverside | US | https://tinyurl.com/yhu6pvm |
| Eidgenössische Technische Hochschule Zürich | CH | https://tinyurl.com/j8u2bcb | University of California, San Diego | US | https://tinyurl.com/j53snwf |
| Emory University | US | https://tinyurl.com/zv42t9x | University of California, San Francisco | US | https://tinyurl.com/znsl2dk |
| George Washington University | US | https://tinyurl.com/j3gcf39 | University of California, Santa Barbara | US | https://tinyurl.com/4ynrs22 |
| Georgetown University | US | https://tinyurl.com/nduolwf | University of Central Florida | US | https://tinyurl.com/hkfyrwd |
| Georgia Institute of Technology | US | https://tinyurl.com/hf9ouxu | University of Chicago | US | https://tinyurl.com/jhk4ylb |
| Georgia State University | US | https://tinyurl.com/zt3nn3q | University of Colorado Boulder | US | https://tinyurl.com/zhxvs3s |
| Iowa State University of Science and Technology | US | https://tinyurl.com/gsr6xgx | University of Connecticut | US | https://tinyurl.com/zvrlmy5 |
| Johns Hopkins University | US | https://tinyurl.com/mzc4a8e | University of Delaware | US | https://tinyurl.com/zsompcs |
| McGill University | CA | https://tinyurl.com/gncj8js | University of Georgia | US | https://tinyurl.com/zbzlejo |
| Monash University | AU | https://tinyurl.com/gncv3pf | University of Houston | US | https://tinyurl.com/hsma5du |
| Nanjing University | CN | https://tinyurl.com/jv964l3 | University of Illinois at Chicago | US | https://tinyurl.com/hpgqm2h |
| National Chiao Tung University | TW | https://tinyurl.com/ju3jpvu | University of Illinois at Urbana-Champaign | US | https://tinyurl.com/hjjxsgq |
| National Tsing Hua University | TW | https://tinyurl.com/j6pl29b | University of Iowa | US | https://tinyurl.com/jl4uqvq |
| National University of Singapore | SG | https://tinyurl.com/htt5kn2 | University of Kansas | US | https://tinyurl.com/j5axnv5 |
| New York University | US | https://tinyurl.com/z7lzjqg | University of Kentucky | US | https://tinyurl.com/lxvprln |
| North Carolina State University | US | https://tinyurl.com/zms93ev | University of Leeds | GB | https://tinyurl.com/jjwnch3 |
| Northwestern University | US | https://tinyurl.com/qajqo6l | University of Maryland | US | https://tinyurl.com/hpsfq59 |
| Oregon State University | US | https://tinyurl.com/jrzuvvy | University of Michigan | US | http://tinyurl.com/jalms7z |
| Penn State University | US | https://tinyurl.com/z46urhn | University of Minnesota | US | https://tinyurl.com/no9orlp |
| Princeton University | US | https://tinyurl.com/hbt5hnn | University of Missouri | US | https://tinyurl.com/nnc82od |
| Purdue University | US | http://tinyurl.com/hry9hca | University of Nebraska-Lincoln | US | https://tinyurl.com/jrbjex3 |
| Rensselaer Polytechnic Institute | US | https://tinyurl.com/z3qjx4n | University of New Mexico | US | https://tinyurl.com/phoxotr |
| Rice University | US | https://tinyurl.com/j9dnca7 | University of North Carolina at Chapel Hill | US | https://tinyurl.com/jo46l4f |
| Rochester Institute of Technology | US | https://tinyurl.com/habtn56 | University of Notre Dame | US | https://tinyurl.com/jlhj6lm |
| Rutgers, The State University of New Jersey | US | https://tinyurl.com/2zk58s | University of Oregon | US | https://tinyurl.com/285fra |
| San Diego State University | US | https://tinyurl.com/hsmrg4b | University of Oxford | GB | https://tinyurl.com/zw6szsr |
| Shanghai Jiao Tong University | CN | https://tinyurl.com/jf2fj2l | University of Pennsylvania | US | https://tinyurl.com/zxqfo9y |
| Stanford University | US | http://tinyurl.com/zvnc9wx | University of Pittsburgh | US | https://tinyurl.com/jy9oeqw |
| Syracuse University | US | https://tinyurl.com/hpg3sr9 | University of Rochester | US | https://tinyurl.com/henjgu6 |
| Texas A&M University | US | https://tinyurl.com/29jwdu2 | University of South Florida | US | https://tinyurl.com/j8mfs54 |
| The Australian National University | AU | https://tinyurl.com/zzb4hxp | University of Southern California | US | https://tinyurl.com/6t5sk8 |
| The University of British Columbia | CA | https://tinyurl.com/hjwbhhc | University of Toronto | CA | https://tinyurl.com/zvusy4d |
| The University of Hong Kong | HK | https://tinyurl.com/hg6ge6z | University of Utah | US | https://tinyurl.com/z3klfhv |
| The University of Manchester | GB | https://tinyurl.com/242wy9 | University of Washington | US | http://tinyurl.com/j8398a4 |
| The University of Melbourne | AU | https://tinyurl.com/hu2b8be | University of Waterloo | CA | https://tinyurl.com/zuxt3qj |
| The Univ. of New South Wales | AU | https://tinyurl.com/zosmkyc | University of Wisconsin-Madison | US | https://tinyurl.com/jmc7o54 |
| The University of Nottingham | GB | https://tinyurl.com/jov5mtr | Vanderbilt University | US | https://tinyurl.com/zathay8 |
| The University of Queensland | AU | https://tinyurl.com/jvbckq6 | Washington State University | US | https://tinyurl.com/gtdr2v7 |
| The University of Tennessee | US | https://tinyurl.com/jkuy39b | Washington Univ. in St. Louis | US | https://tinyurl.com/a5g7u27 |
| The Univ. of Texas at Austin | US | https://tinyurl.com/5nehcf | Yale University | US | https://tinyurl.com/zdrypbu |